\documentclass[9pt,twocolumn,twoside]{osajnl}
\usepackage{comment}
\usepackage{amsmath}
\usepackage{lipsum}
\usepackage{float}
\usepackage{nicefrac}
\usepackage{soul}
\usepackage{cancel}
\usepackage[normalem]{ulem}

\journal{ol} 

\setboolean{shortarticle}{true}
\usepackage{multirow,tabularx}

\title{Mode-selective Image Upconversion}

\author[1,2]{Santosh Kumar}
\author[1,2]{He Zhang}
\author[1,2]{Stephanie Maruca}
\author[1,2,*]{Yu-Ping Huang}
\affil[1]{Department of Physics, Stevens Institute of Technology, Hoboken, NJ, 07030, USA}
\affil[2]{Center for Quantum Science and Engineering, Stevens Institute of Technology, Hoboken, NJ, 07030, USA}

\affil[*]{yuping.huang@stevens.edu}

 \dates{Compiled \today}

\ociscodes{ (190.0190) Nonlinear optics, (190.7220) Upconversion, (280.3640), (030.1670) Coherent optical effects, (070.6120) Spatial light modulators,
(030.4070) Modes.}

\begin{abstract}
We study selective upconversion of optical signals according to their detailed transverse electromagnetic modes, and demonstrate its proof of operations in a nonlinear crystal. The mode selectivity is achieved by preparing the pump wave in an optimized spatial profile to drive the upconversion. For signals in the Laguerre-Gaussian modes, we show that a mode can be converted with up to 60 times higher efficiency than an overlapping but orthogonal mode. This nonlinear-optical approach may find applications in compressive imaging, pattern recognition, quantum communications, and others where the existing linear-optical methods are limited. 
\end{abstract}

\setboolean{displaycopyright}{true}

\begin{document}

\maketitle
 

Optical parametric conversion is an instrumental resource in classical and quantum optics \cite{Boyd03,Kumar90,Lassen07,Steinlechner16,Chaitanya16,Burenkov17}. In one of its latest developments, mode-selective conversion (MSC) has emerged as a versatile tool capable of discriminatively converting overlapping optical signals according to their orthogonal electromagnetic modes for various applications \cite{Benjamin11,Eckstein11,Reddy13,YuPing13,Brecht13}. To this end, there have been rapid experimental progresses made in the time-frequency domain using nonlinear waveguides, with MSC demonstrated among picosecond pulses and their superposition states \cite{Kowligy14,Brecht14,Manurkar16,Reddy17,Allgaier17}, and for a single mode over broadband noise \cite{QPMS2017,Amin18}. Those results herald some unprecedented applications in quantum communications, quantum computing, quantum LiDAR, remote sensing, and so on. 

In this Letter, we extend those exciting studies to the spatial domain, where similar benefits are expected for a wealth of free-space applications. Rather than using the waveguides and time-frequency modes, here we pump a second-order nonlinear crystal with a laser in a tailored transverse mode, thereby achieving selective conversion of signals in overlapping but orthogonal spatial modes with largely differential efficiencies. Our experiment highlights a nonlinear-optical approach to generation, manipulation, and detection of higher-order spatial modes and structured optical images, with potential applications in photon-efficient quantum communications, object rectification, compressive sensing, pattern recognition, etc.
\cite{Mesaritakis15,Aguiar16,Berkhout10,Bozinovic13,Mirhosseini13,Labroille14,Huang15,Wittek16}. Comparing with their linear-optical counterparts, 
this and other nonlinear-optics techniques could offer practical advantages \cite{Delaubert07,Lanning17,Pereira17,Demur18}, such as those  demonstrated in spiral phase contrast imaging of the edges \cite{Xiaodong18} and the enhancement of the field-of-view \cite{Maestre18}. In particular, the recent studies of MSC in multimode waveguides direct to high-speed quantum communications over few-mode optical fibers \cite{Vasilyev14,Vasilyev17}.     %

Our spatial MSC is implemented through sum-frequency (SF) generation in a periodic-poled lithium niobate (PPLN) crystal. The mode selectivity is realized by preparing the pump in an optimized spatial mode via adaptive feedback control. The pump and signal modes are each prepared by using spatial light modulators (SLMs). Specifically, we consider the signal in the Laguerre-Gaussian (LG) modes with helical wavefronts, which are written in the cylindrical coordinates as
\begin{multline}
\psi \equiv LG_l^{p}(r,\phi,z)=\frac{C_{lp}}{\omega(z)}\left(\frac{r\sqrt{2}}{\omega(z)}\right)^{|l|}\exp\left(\frac{-r^{2}}{\omega^{2}(z)}\right) \exp(-ikz)\\ \times L_{p}^{|l|}\left(\frac{2r^{2}}{\omega^{2}(z)}\right) \exp\left(-ik\frac{r^{2}}{2R(z)}\right)\exp(-il\phi)\exp(i\zeta(z)).
\end{multline}
Here, $\{\displaystyle L_{p}^{\vert l \vert}\}$ are the generalized Laguerre polynomials with the azimuthal mode index $l$ and the radial index $p$, ${\displaystyle C_{lp}}=\sqrt{\frac{2p!}{\pi(p+|l|)!}}$ is a normalization constant, ${\displaystyle w(z)}$ =$w_0$ 
$\sqrt{1 + (z/z_R)^2}$ is the beam radius at $z$, $w_0$ is the beam waist, $z_R$ = $\pi w_0^2/\lambda$ is the Rayleigh range, ${\displaystyle R(z)}$ = $z$
$(1 + (z_R/z)^2)$ is the radius of curvature of the beam, $\kappa$ = $2\pi n/\lambda$ is the wave number, and $\zeta(z)$ is the Gouy phase at $z$. Here, we demonstrate the MSC of the LG modes that are actively deployed for classical and quantum communications. The technique itself, however, is applicable to other mode basis, including Hermite-Gauss (HG) modes.

\begin{figure*}[ht]
\centering 
\includegraphics[width=\linewidth]{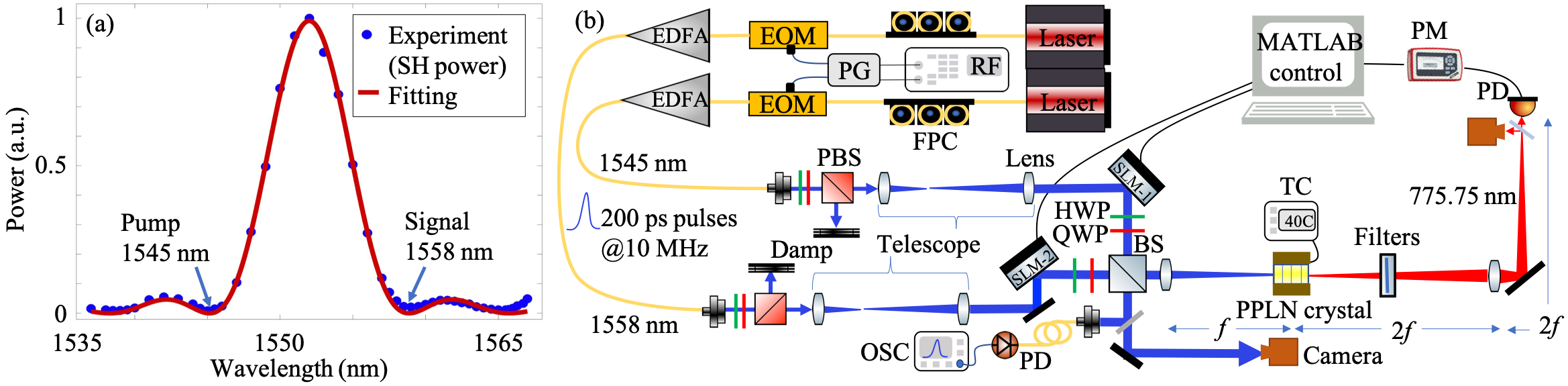}
\caption{(a) Quasi-phase matching profile of the PPLN nonlinear crystal, where the blue dots show the experiment results and the red line shows the $sinc^2$ fitting. The FWHM phase matching bandwidth is about 6.5 nm centered at 1552 nm. (b) Experimental setup for spatial mode-selective conversion. Two synchronized light pulse trains at wavelength 1558 nm and 1545 nm with 200 ps pulse width and 10 MHz repetition rate, pass through a temperature stabilized PPLN crystal. The SLMs are used to create the desired spatial modes. The generated SF modes at wavelength 775.75 nm are filtered and detected using a power meter or CCD camera. The detected SF power feeds into a MATLAB code to adaptively optimize the phase mask on SLM-2. FPC: Fiber Polarization Controller, EOM: Electro-optic Modulator, EDFA: Erbium-doped Fiber Amplifier, QWP: Quarter Waveplate, HWP: Half Waveplate, OSC: Oscilloscope, BS: Beamsplitter, SLM: Spatial Light Modulator, PPLN crystal: Magnesium-doped Periodic Poled Lithium Niobate crystal, PG: Pulse Generator, RF: Radio-Frequency Source, PD: Photodiode, PM: Powermeter, TC: Temperature controller. 
} \label{ExpSetUp}
\end{figure*}

At present, we focus on the SF generation in the non-diffraction regime, where the diffraction of the interacting waves through the crystal is not significant (diffraction would be exploited in the future studies to achieve better performance, similar to our previous studies in the time-frequency domain \cite{QPMS2017}). In this case, the SF generation is described under the slowly-varying-envelope approximation as: 
\begin{eqnarray}
\partial_z\psi_{s}&=&i \left(\frac{\omega_{s}}{n_{s} c}\right) \chi^{(2)} \psi_{p}^{*} \psi_{f} e^{-i\Delta\kappa z}, \\
\partial_z\psi_{f}&=&i \left(\frac{\omega_{f}}{n_{f} c}\right) \chi^{(2)} \psi_{p} \psi_{s} e^{i\Delta\kappa z},
\end{eqnarray}
where $\Delta\kappa=\kappa_{s}+\kappa_{p}-\kappa_{f}-2\pi/\Lambda$ is the momentum mismatching and $\Lambda$ is the poling period of the nonlinear crystal. The subscript $j = {s,p,f}$ denotes the signal, pump, and SF light, 
respectively. The refractive index for the signal, pump, and SF lights are $n_p$, $n_s$, and $n_f$, respectively. $\psi_{p}$ is a constant in the non-depletion pump approximation, which is the case in this study. Under the quasi-phase matching (QPM) condition ($\Delta\kappa\approx0$),  the generated SF light is simply 
\begin{equation}
\psi_{f}(z)=A\psi_{s}(0)\sin \left[\frac{\omega_{s}\omega_{f}}{c^{2}\sqrt{k_{s}k_{f}}}\psi_{p} \chi^{(2)} z\right],
\end{equation}
where $A \equiv i\sqrt{\frac{\omega_{f}n_{s}}{\omega_{s}n_{f}}}$ and $c$ is the speed of light in vacuum. This equation is an analytic result for any point ($x,y$) in the transverse plane of the light waves, which is valid in the current parameter regime without strong diffraction. The mode conversion can then be calculated via integration over the ($x,y$) plane.

In Figure \ref{ExpSetUp}(a), we measure the QPM curve for the PPLN crystal by scanning a continuous-wave laser and measuring the generated second harmonic (SH) power, and fit it with $sinc^{2}$ phase matching function. This result guides us to select the wavelength of the signal and pump beams for SF generation while ensures the minimum production of the second harmonic generation.
Figure \ref{ExpSetUp}(b) gives the schematic of the experimental setup. Two lasers (LaserBlade, Coherent Solutions) generate continuous-wave lightwaves, each at 1558 nm and 1545 nm, with $\leq$ 100 kHz linewidth. The wavelength difference is about 13 nm, so their group velocities are approximately equal in the crystal. These two beams pass through two electro-optic modulators to create two pulse trains, each with 200-ps full width at half maximum (FWHM) and a 10-MHz repetition rate. The resulting pulses are synchronized with a common radio-frequency reference source. To obtain high peak power, the resulting pulses are then amplified using two Erbium-Doped Fiber Amplifiers (EDFAs). The pump's average power is 40 mW and peak power is 21 W. The signal’s average power is 35 mW and peak power is 18 W. After passing through beam expanders, the signal and pump (beam FWHM: 2.6 mm and 2.8 mm, respectively) are then guided through free space to incident on 1.5 cm$\times$1.1 cm SLMs (Santec SLM-100) \cite{Maruca18}. The angle of incidence for the pump and signal beams onto the SLMs are about $50^{\circ}$ and $55^{\circ}$, respectively. 
The SLMs are used to create the desired spatial modes using numerically generated phase mask by discretizing the phase of the $LG_{l}^{p}$ mode in the Fourier domain over a 1 cm $\times$ 1 cm area (smaller than the SLM screen) with 10.4-$\mu$m pixel pitch, which is $\varphi(r,\phi) = -l \phi+\pi \theta\left(-L_{p}^{|l|}\left(\frac{2r^{2}}{\omega_{0}^{2}}\right)\right)$ ($\theta$ is a unit step function) \cite{Matsumoto2008}. The two pulse trains are then combined at a beam splitter (BS) and focused (focus length, f=200 mm) inside a temperature-stabilized PPLN crystal, quasi phase matched, with a poling period of 19.36 $\mu$m (5 mol.\% MgO doped PPLN, 3 mm width, 2 mm length, and 1 mm height) for frequency conversion to produce the SF light at 775.75 nm. The light, which is polarized vertically along the y-axis (parallel to the crystal’s optical axis), 
propagates through the crystal along the z-direction. The beam waist of the Gaussian signal and Gaussian pump inside the crystal are 45 $\mu$m and 41 $\mu$m, respectively. The output pulses are then filtered with 2 short-pass filters to select the SF and remove any residual light \cite{YMS, QPMS2017}. The other arm of the BS is used to monitor the spatial mode of the pump and signal beams using a near-infrared camera (FIND-R-SCOPE Model No. 85700) with a 17.6 $\mu$m pixel resolution. The same arm is also used to temporally align the two pulses and monitor intensity fluctuation on an oscilloscope (HP 54750A). The SF light is captured after a lens imaging system using a CCD camera (Canon Rebel T5) with a sensor size of 22.3 mm $\times$ 14.9 mm and a pixel pitch of 4.3 $\mu$m. The depth of focus is 5 mm with an Airy disk diameter of 7 $\mu$m (1.22 $\lambda f_N$, where $f_N$ is the f-number of a lens). 

\begin{figure}[htbp]
  \centering
   \includegraphics[width=7.0cm]{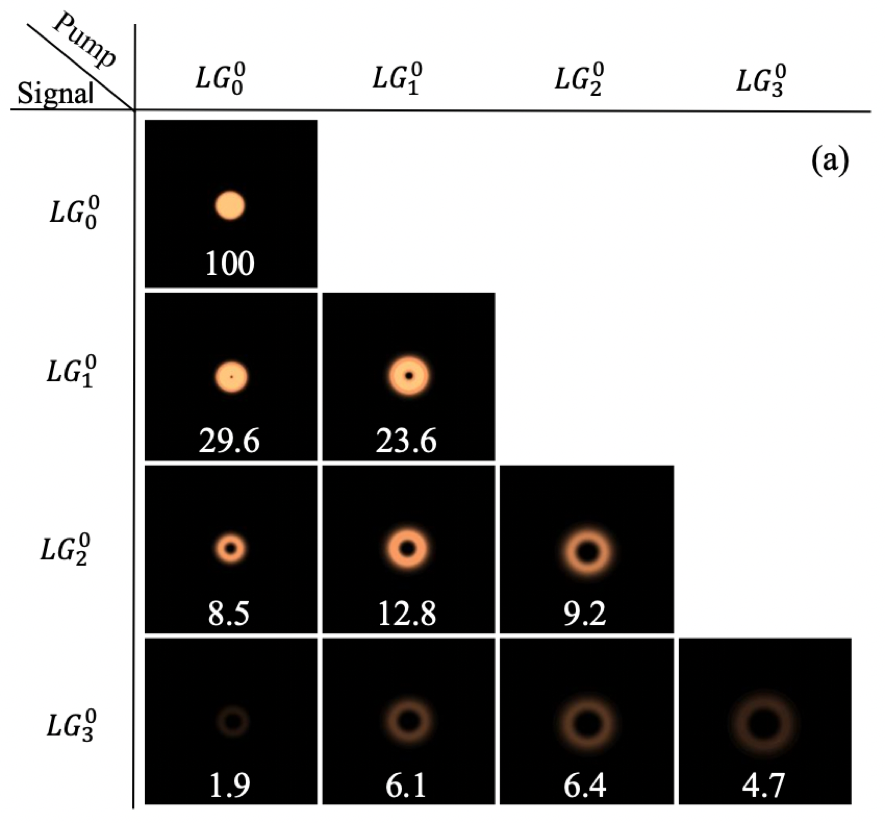}
   \includegraphics[width=7.0cm]{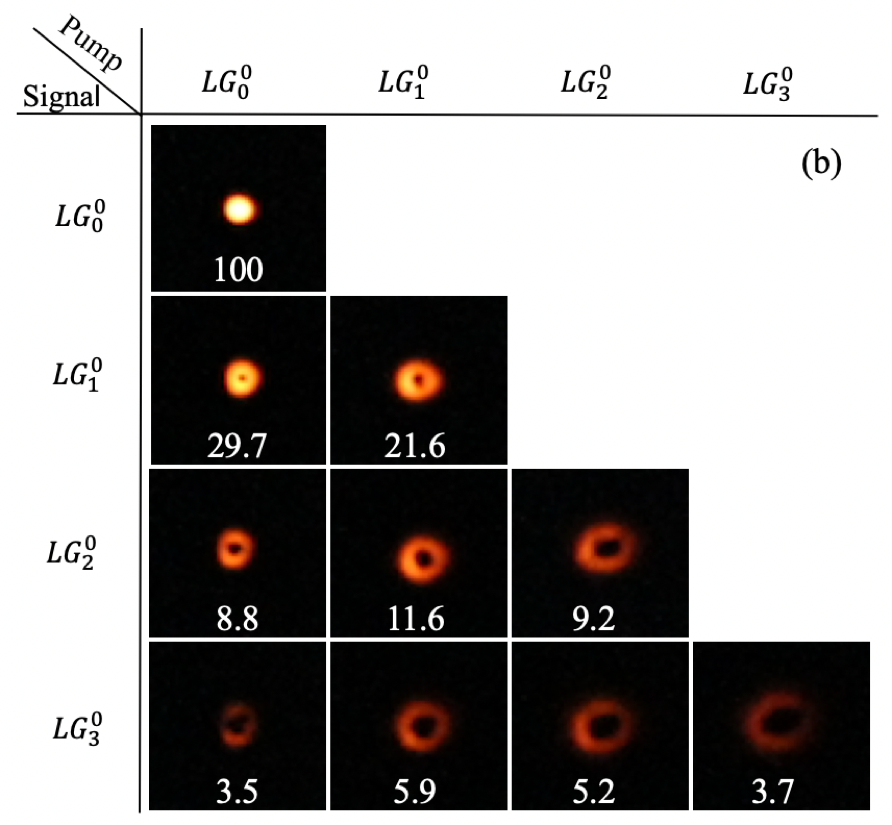} 
\caption{(a) Simulated and (b) experimental results of generated SF spatial $LG_l^p$ modes. 
The values inside the figure of each mode are the percent relative conversion efficiencies, $\eta$\%.
} \label{modes}
\end{figure}

Restricted by the available laser power and a small length of the crystal (2 mm), the present measurements are in the low conversion regime. Therefore, rather than using the absolute conversion efficiency, we define $\eta$ as the relative conversion efficiency, which is the efficiency relative to that of a Gaussian pump with a Gaussian signal of the same beam waist and same total pulse energy. Then, the conversion extinction between two signal modes, $\xi$, is defined as the ratio of their conversion efficiencies by the same pump. In Fig. \ref{modes} (a) and (b), we show the experimental and numerical results of the SF spatial profiles created by different combinations of $LG^p_{l}$ modes for the signal and pump, respectively, with $\eta$ for each combination. These results show some clear dependence of the conversion on the pump and signal modes. However, there is no  appreciable extinction as the pump modes have yet to be optimized. For example, only a modest selectivity of 5 dB is achieved between two signal modes $LG^0_{1}$ and $LG_{0}^0$ with the same pump mode $LG_0^0$. 

To achieve high selectivity, we use a particle swarm optimization (PSO) algorithm to numerically optimize the pump spatial modes for targeted signal modes \cite{Kennedy95}. In this method, a particle corresponds to a set of weighted phase masks for LG modes, whose superposition produces the pump profile after the Fourier lens. The personal and global bests correspond to the highest selectivities achieved by each individual particle and by all particles, respectively \cite{Saremi07,Clerc10}. The optimum phase mask for pump mode is obtained after reaching a desirable selectivity as shown in Table \ref{Opt_modes} (a). We also compare the analytic results obtained under the undepleted pump approximation and exact numeric results taking into count the pump depletion and found good agreement.

\begin{figure}[htbp]
  \centering
  \includegraphics[width=\linewidth]{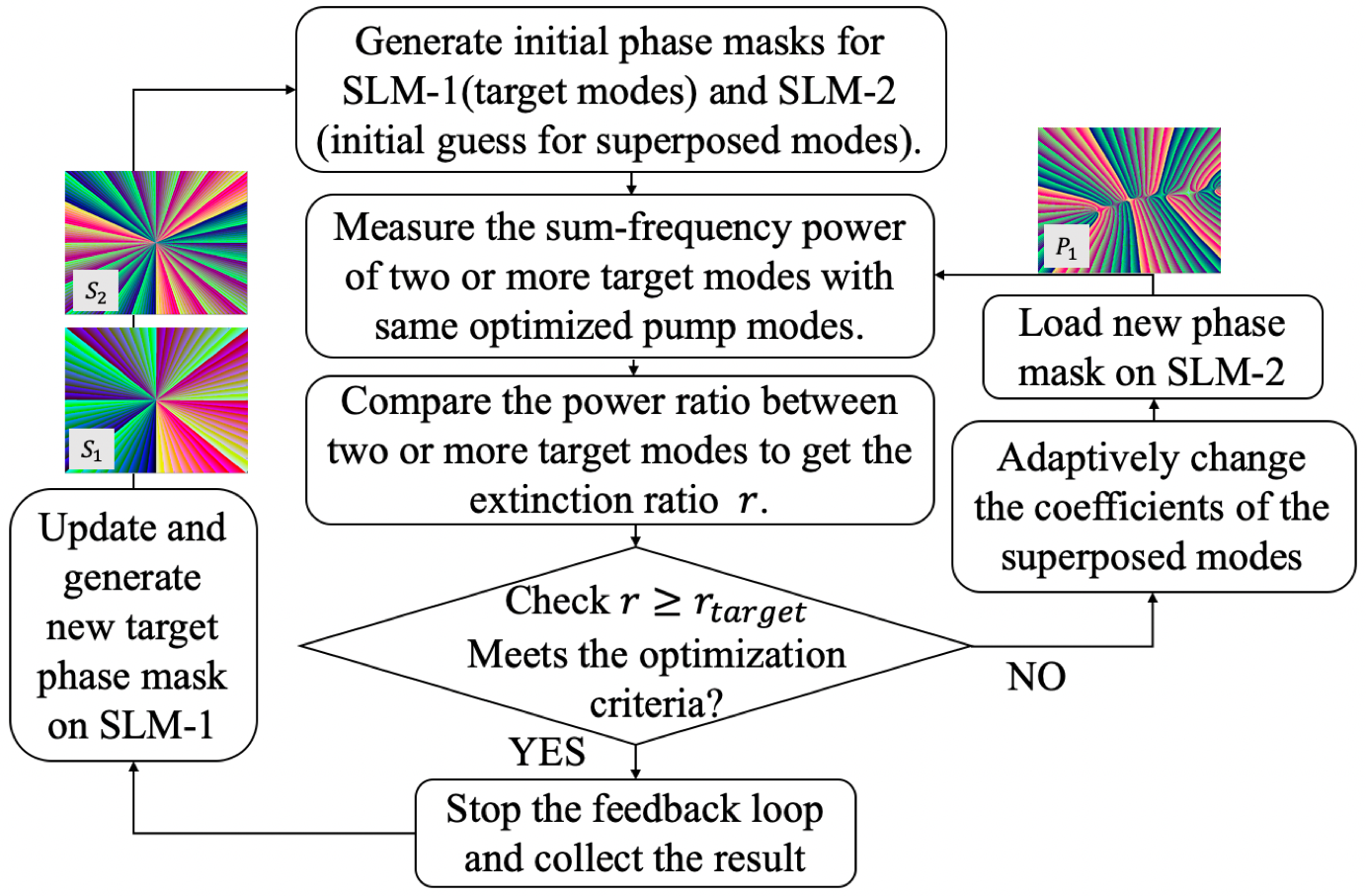}
  \caption{Flow chart of the adaptive feedback control algorithm. $S_1$ and $S_2$ are the phase masks for the signal modes, $P_1$ is the optimized phase mask for the pump mode. The phase masks are generated using 10-bit encoding format in RGB color for Sentec-SLM. {\it Step 1:} Generate initial phase masks for SLM-1 and SLM-2. {\it Step 2:} Measure the up-converted power of two or more target modes applied on the SLM-1 with the same superposed modes on SLM-2. {\it Step 3:} Compute the extinction ratio, $r$, between the generated modes. {\it Step 4:} Check if the optimization criteria, $r \geq r_{target}$, is met. {\it Step 5:} If  $r < r_{target}$, generate the new superposition coefficients for the mode basis and update the phase mask for SLM-2. {\it Step 6:} Repeat {\it Steps 2-5} until  $r \geq r_{target}$. {\it Step 7:} Collect the optimum selectivity between the desired modes. {\it Step 8:} Update the target phase mask on SLM-1 and repeat {\it Steps 1-7}. }
  \label{Flowchart}
\end{figure}

In experiment, we apply the numerically optimized phase masks to create the pump mode. However, due to the phase errors in the SLM and the imperfect alignment of the Gaussian beam with respect to the phase mask, the attained selectivity is reduced. To account for such imperfections, we utilize adaptive feedback control in the experiment \cite{Tzang18}, where two SLMs (SLM-1 and SLM-2) and a power-meter (Thorlabs PM-100D with sensor S130C) are automated through a computer using a MATLAB interface. A Flow chart of the adaptive feedback control algorithm is shown in Fig. \ref{Flowchart}. The phase masks are loaded onto the SLM-1 and SLM-2 for the signal and pump, respectively. In this feedback control process, we adaptively manipulate phase mask for the SLM-2 to optimize the selectivity between the two signal modes.

Table \ref{Opt_modes} shows the (a) simulation and (b) experimental results for the optimized MSC by tailoring the phase masks for the pump. Our goal is to selectively up-convert one particular signal mode against other overlapping signal modes, or vice versa. To quantify the performance, $\xi$ is the ratio of the conversion efficiencies for the target mode against the undesirable mode. We consider three examples here. In the first case, the Pump $P^+_{00}$ is designed to convert the $LG^0_{0}$ mode while simultaneously suppressing LG$_1^0$, LG$_2^0$, and LG$_3^0$ modes, respectively. The simulation achieves an extinction of 6.5 dB, 13.1 dB, and 20.9 dB for each, while the experimental results are 5.9 dB, 12.1 dB, and 18.1 dB. The second case is the exact opposite, where Pump $P^-_{00}$ is designed to suppress LG$_0^0$, but convert LG$_1^0$, LG$_2^0$, and LG$_3^0$. The numerically maximized extinctions are 16.5 dB, 20.4 dB, and 21.7 dB for each, while the experimental results are 11.8 dB, 14.8 dB, and 15.6 dB, respectively. In the last example, Pump $P^+_{01}$ is optimized to convert $LG^0_{1}$ against $LG^0_{2}$, $LG^0_{3}$, and $LG^0_{4}$, respectively. The simulation gives 6.6 dB, 14.5 dB, and 22.4 dB extinction for each, while the experiment produces 5.6 dB, 11.5 dB, and 15.2 dB. From the table, the experimental measurements are in good agreement with the numerical simulations without using any fitting parameter, which validates our approach. In the future, we will extend this method to the high conversion regime, where higher selectivities are expected, as evident in the time-frequency domain \cite{QPMS2017}. 

\begin{table}[htbp]
 \centering
\arrayrulecolor{gray}

  \centering
\begin{tabular}{|c|c|c|c|c|c|c|}

\hline 
 \rowcolor[gray]{0.9} {\bf (a)} &  & $\xi$ (dB) &  & $\xi$ (dB) & & $\xi$ (dB) \tabularnewline
  \hline 
\cellcolor[gray]{0.9} & $LG_{0}^{0}$ & \multirow{2}{*}{6.5} & $LG_{0}^{0}$ & \multirow{2}{*}{13.1} & $LG_{0}^{0}$ & \multirow{2}{*}{20.9}\tabularnewline
\cline{2-2} \cline{4-4} \cline{6-6} 
 \multirow{-2}{*}{\cellcolor[gray]{0.9}$P^+_{00}$} & $LG_{1}^{0}$ &   & $LG_{2}^{0}$ &  & $LG_{3}^{0}$ &  \tabularnewline
\hline 
\cellcolor[gray]{0.9} & $LG_{1}^{0}$ &  \multirow{2}{*}{16.5} & $LG_{2}^{0}$ & \multirow{2}{*}{20.4} & $LG_{3}^{0}$ & \multirow{2}{*}{21.7}\tabularnewline
\cline{2-2} \cline{4-4} \cline{6-6} 
\multirow{-2}{*}{\cellcolor[gray]{0.9}$P^-_{00}$} & $LG_{0}^{0}$ &  & $LG_{0}^{0}$ &  & $LG_{0}^{0}$ & \tabularnewline
\hline 
\cellcolor[gray]{0.9} & $LG_{1}^{0}$ &\multirow{2}{*}{6.6} & $LG_{1}^{0}$ & \multirow{2}{*}{14.5} & $LG_{1}^{0}$ & \multirow{2}{*}{22.4}\tabularnewline
\cline{2-2} \cline{4-4} \cline{6-6} 
 \multirow{-2}{*}{\cellcolor[gray]{0.9}$P^+_{01}$} & $LG_{2}^{0}$ &  & $LG_{3}^{0}$ &  & $LG_{4}^{0}$ &  \tabularnewline
\hline

\end{tabular}%

\begin{tabular}{|c|c|c|c|c|c|c|}

 \hline 
 \rowcolor[gray]{0.9} {\bf (b)} &  & $\xi$ (dB) &  & $\xi$ (dB) & & $\xi$ (dB) \tabularnewline
 \hline 
\cellcolor[gray]{0.9} & $LG_{0}^{0}$ &  \multirow{2}{*}{5.9} & $LG_{0}^{0}$ &  \multirow{2}{*}{12.1} & $LG_{0}^{0}$ &  \multirow{2}{*}{18.1}\tabularnewline
\cline{2-2} \cline{4-4} \cline{6-6} 
 \multirow{-2}{*}{\cellcolor[gray]{0.9}$P^+_{00}$} & $LG_{1}^{0}$ &   & $LG_{2}^{0}$  &   & $LG_{3}^{0}$ & \tabularnewline
\hline 
\cellcolor[gray]{0.9} & $LG_{1}^{0}$ & \multirow{2}{*}{11.8} & $LG_{2}^{0}$ &  \multirow{2}{*}{14.8} & $LG_{3}^{0}$ &  \multirow{2}{*}{15.6}\tabularnewline
\cline{2-2} \cline{4-4} \cline{6-6} 
  \multirow{-2}{*}{\cellcolor[gray]{0.9}$P^-_{00}$}& $LG_{0}^{0}$ &   & $LG_{0}^{0}$ &  & $LG_{0}^{0}$ &  \tabularnewline
\hline 
\cellcolor[gray]{0.9} & $LG_{1}^{0}$ & \multirow{2}{*}{5.6} & $LG_{1}^{0}$ &  \multirow{2}{*}{11.5} & $LG_{1}^{0}$ & \multirow{2}{*}{15.2}\tabularnewline
\cline{2-2} \cline{4-4} \cline{6-6} 
  \multirow{-2}{*}{\cellcolor[gray]{0.9}$P^+_{01}$}& $LG_{2}^{0}$ &   & $LG_{3}^{0}$ &   & $LG_{4}^{0}$ &  \tabularnewline
\hline 

\end{tabular}%

 
\caption{ (a) Simulation and (b) experimental results for the spatial MSC using pump modes in optimized superpositions of LG modes. 
} \label{Opt_modes}
\end{table}

In summary, we have demonstrated the selective upconversion of overlapping spatial modes in a nonlinear crystal by optimizing the pump profiles. Our experimental results, which achieve conversion extinction up to 18 dB, are in good agreement with the numerical simulation without the use of any fitting parameters. The next step will be to extend this technique to the high conversion regime with strong diffraction, where even better performance is expected. The nonlinear optics based approach may find novel applications such as improving the signal-to-noise ratio to recognize an image pattern \cite{Aguiar16,Mesaritakis15}, spatial mode de-multiplexing of higher-order modes for classical and quantum communications \cite{Vasilyev14,Vasilyev17}, and mode-selective LiDAR \cite{Amin18}. 

This research was supported in part by the Office of Naval Research (Award No. N00014-15-1-2393). The authors thanks Yong Meng Sua for useful discussions.

\newpage
\section{Full References}


\begin{thebibliography}{1}

\bibitem{Kumar90} P. Kumar, ``Quantum frequency conversion," 
Opt. Lett. \textbf{15}, 1476-1478 (1990).

\bibitem{Boyd03} R. Boyd, {\it Nonlinear Optics} (Academic Press, USA, 2003).

\bibitem{Lassen07} M. Lassen, V. Delaubert, J. Janousek, K. Wagner, H.-A. Bachor, P.K. Lam, N. Treps, P. Buchhave, C. Fabre and C. C. Harb, ``Tools for Multimode Quantum Information: Modulation, Detection, and Spatial Quantum Correlations," 
Phys. Rev. Lett. \textbf{98}, 083602 (2007).

\bibitem{Steinlechner16} F. Steinlechner, N. Hermosa, V. Pruneri and J. P. Torres, ``Frequency conversion of structured light," 
Sci. Rep. \textbf{6}, 21390 (2016).

\bibitem{Chaitanya16} N.A. Chaitanya, M.V. Jabir, J. Banerji and G.K. Samanta, ``Hollow Gaussian beam generation through nonlinear interaction of photons with orbital angular momentum," 
Sci. Rep. \textbf{6}, 32464 (2016).

\bibitem{Burenkov17} I. A. Burenkov, T. Gerrits, A. Lita, S. W. Nam, L. Krister Shalm, and S. V. Polyakov, ``Quantum frequency bridge: high-accuracy characterization of a nearly-noiseless parametric frequency converter," 
Opt. Express \textbf{25}, 907-917 (2017).

\bibitem{Benjamin11} B. Benjamin, A. Eckstein, A. Christ, H. Suche and C. Silberhorn, ``From Quantum Pulse Gate to Quantum Pulse Shaper-Engineered Frequency Conversion in Nonlinear Optical Waveguides.” 
New Jour. of Phys. \textbf{13}, 065029 (2011).

\bibitem{Eckstein11} A. Eckstein, B. Brecht, and C. Silberhorn, ``A quantum pulse gate based on spectrally engineered sum frequency generation," 
Opt. Express \textbf{19}, 13770-13778 (2011).

\bibitem{Reddy13} D. V. Reddy, M. G. Raymer, C. J. McKinstrie, L. Mejling, and K. Rottwitt, ``Temporal mode selectivity by frequency conversion in second-order nonlinear optical waveguides," 
Opt. Express \textbf{21}, 13840-13863 (2013).

\bibitem{YuPing13} Y.-P. Huang and P. Kumar, ``Mode-resolved photon counting via cascaded quantum frequency conversion," 
Opt. Lett. \textbf{38}, 468-470 (2013).

\bibitem{Brecht13}  B. Brecht, D. Reddy, C. Silberhorn and M. Raymer, ``Photon Temporal Modes: A Complete Framework for Quantum Information Science," 
Phys. Rev. X \textbf{5}, 041017 (2015).

\bibitem{Allgaier17} M. Allgaier, V. Ansari, L. Sansoni, C. Eigner, V. Quiring, R. Ricken, G. Harder, B. Brecht, and C. Silberhorn, ``Highly efficient frequency conversion with bandwidth compression of quantum light," 
Nat. Commun. \textbf{8}, 14288 (2017).

\bibitem{Kowligy14} A. S. Kowligy, P. Manurkar, N. V. Corzo, V. G. Velev, M. Silver, R. P. Scott, S. J. B. Yoo, P. Kumar, G. S. Kanter, and Y.-P. Huang, ``Quantum optical arbitrary waveform manipulation and measurement in real time," 
Opt. Express \textbf{22}, 27942-27957 (2014)


\bibitem{Brecht14} B. Brecht, A. Eckstein, R. Ricken, V. Quiring, H. Suche, L. Sansoni, and C. Silberhorn, ``Demonstration of coherent time-frequency Schmidt mode selection using dispersion-engineered frequency conversion", 
Phys. Rev. A \textbf{90}, 030302 (2014).

\bibitem{Manurkar16} P. Manurkar, N. Jain, M. Silver, Y.-P. Huang, C. Langrock, M. M. Fejer, P. Kumar, and G. S. Kanter, ``Multidimensional mode-separable frequency conversion for high-speed quantum communication," 
Optica \textbf{3}, 1300-1307 (2016).

\bibitem{Reddy17} D. V. Reddy and M. G. Raymer, ``Engineering temporal-mode-selective frequency conversion in nonlinear optical waveguides: from theory to experiment," 
Opt. Express \textbf{25}, 12952-12966 (2017).

\bibitem{QPMS2017} A. Shahverdi A, Y. M. Sua , L. Tumeh  and Y-P. Huang, ``Quantum Parametric Mode Sorting: Beating the Time-Frequency Filtering," 
Sci. Rep. \textbf{7}, 6495 (2017).

\bibitem{Amin18} A. Shahverdi, Y. M. Sua, I. Dickson, M. Garikapati, and Y-P. Huang, ``Mode selective up-conversion detection for LIDAR applications," 
Opt. Express \textbf{26}, 15914-15923 (2018).

\bibitem{Mesaritakis15} C. Mesaritakis, A. Bogris, A. Kapsalis, and D. Syvridis, ``High-speed all-optical pattern recognition of dispersive Fourier images through a photonic reservoir computing subsystem," 
Opt. Lett. \textbf{40}, 3416-3419 (2015).

\bibitem{Aguiar16} H. B. de Aguiar, S. Gigan, and S. Brasselet, ``Enhanced nonlinear imaging through scattering media using transmission-matrix-based wave-front shaping," 
Phys. Rev. A \textbf{94}, 043830 (2016).

\bibitem{Berkhout10} G. C. Berkhout, M. P. Lavery, J. Courtial, M. W. Beijersbergen, M. J. Padgett, ``Efficient sorting of orbital angular momentum states of light,"
Phys. Rev. Lett. \textbf{105}, 153601 (2010).

 \bibitem{Bozinovic13} N. Bozinovic, Y. Yue, Y. Ren, M. Tur, P. Kristensen, H. Huang, A.E. Willner and S. Ramachandran, ``Terabit-Scale Orbital Angular Momentum Mode Division Multiplexing in Fibers, 
 Science \textbf{340}, 1545-1548 (2013).

\bibitem{Mirhosseini13} M. Mirhosseini,  M. Malik, Z. Shi, R. Boyd, ``Efficient separation of the orbital angular momentum eigenstates of light,"
Nat. Commun. \textbf{4}, 2781 (2013).
 
\bibitem{Labroille14} G. Labroille, B. Denolle, P. Jian, P.e Genevaux, N. Treps, and J.-F. Morizur, ``Efficient and mode selective spatial mode multiplexer based on multi-plane light conversion," 
Opt. Express \textbf{22}, 15599-15607 (2014).

 \bibitem{Huang15} H. Huang, G. Milione, M.P.J. Lavery, G. Xie, Y. Ren, Y. Cao, N. Ahmed, T.A. Nguyen, D.A. Nolan, M.-J. Li, M. Tur, R.R. Alfano and A.E. Willner, ``Mode division multiplexing using an orbital angular momentum mode sorter and MIMO-DSP over a graded-index few-mode optical fibre, 
 Sci. Rep. \textbf{5}, 14931 (2015).

\bibitem{Wittek16} S. Wittek, R.B. Ramirez, J.A. Zacarias, Z.S. Eznaveh, J. Bradford, G.L. Galmiche, D. Zhang, W. Zhu, J.A.-Lopez, L. Shah, and R.A. Correa, ``Mode-selective amplification in a large mode area Yb-doped fiber using a photonic lantern," 
Opt. Lett. \textbf{41}, 2157-2160 (2016).

\bibitem{Delaubert07} V. Delaubert, M. Lassen, D.R.N. Pulford, H-A. Bachor, and C.C. Harb, ``Spatial mode discrimination using second harmonic generation," 
Opt. Express \textbf{15}, 5815-5826 (2007).

\bibitem{Lanning17} R.N. Lanning, Z. Xiao, M. Zhang, I. Novikova, E.E. Mikhailov, and J.P. Dowling, ``Gaussian-beam-propagation theory for nonlinear optics involving an analytical treatment of orbital-angular-momentum transfer," 
Phys. Rev. A \textbf{96}, 013830 (2017).

\bibitem{Pereira17} L.J. Pereira, W.T. Buono, D.S. Tasca, K. Dechoum, and A.Z. Khoury, ``Orbital-angular-momentum mixing in type-II second-harmonic generation,"
Phys. Rev. A \textbf{96}, 053856 (2017).

\bibitem{Demur18} R. Demur, R. Garioud, A. Grisard, E. Lallier, L. Leviandier, L. Morvan, N. Treps, and C. Fabre, ``Near-infrared to visible upconversion imaging using a broadband pump laser," 
Opt. Express \textbf{26}, 13252-13263 (2018).

\bibitem{Xiaodong18} X. Qiu, F. Li, W. Zhang, Z. Zhu, and L. Chen, ``Spiral phase contrast imaging in nonlinear optics: seeing phase objects using invisible illumination," 
Optica \textbf{5}, 208-212 (2018).

\bibitem{Maestre18} H. Maestre, A. J. Torregrosa, C. R. Fernández-Pousa, and J. Capmany, ``IR-to-visible image upconverter under nonlinear crystal thermal gradient operation," 
Opt. Express \textbf{26}, 1133-1144 (2018).

\bibitem{Vasilyev14} M. Vasilyev, Y.B. Kwon, and Y. Huang, ``Spatial-Mode-Selective Quantum Frequency Conversion in a $\chi^{(2)}$ Slab Waveguide," 
OSA Technical Digest (online), paper JW2A.52 (2014).

\bibitem{Vasilyev17} Y.B. Kwon, M. Giribabu, C. Langrock, M.M. Fejer, and M. Vasilyev, ``Single-photon-level spatial-mode-selective frequency up-conversion in a multimode $\chi^{(2)}$ waveguide," 
OSA Technical Digest (online), paper FF2E.2 (2017).

\bibitem{Maruca18} S. Maruca, S. Kumar, Y.M. Sua, J.-Y. Chen, A. Shaherdi, and Y.-P. Huang, ``Quantum Airy Photons," 
J. Phys. B: At. Mol. opt. Phys. \textbf{51},  175501 (2018).

\bibitem{Matsumoto2008} N. Matsumoto, T. Ando, T. Inoue, Y. Ohtake, N. Fukuchi, and T. Hara, 
``Generation of high-quality higher-order Laguerre-Gaussian beams using liquid-crystal-on-silicon spatial light modulators," 
 J. Opt. Soc. Am. A \textbf{25}, 1642-1651 (2008).

\bibitem{YMS}  Y.M. Sua, H. Fan, A. Shahverdi, J. Chen and Y-P. Huang, ``Direct Generation and Detection of Quantum Correlated Photons with 3.2 um Wavelength Spacing," 
Sci. Rep. \textbf{7}, 17494 (2017).

\bibitem{Kennedy95} J. Kennedy and  R. Eberhart, ``Particle swarm optimization," Proc. IEEE Int. Conf. Neural Netw. \textbf{4} 1942, (1995).

\bibitem{Saremi07} M. S.-Saremi and R. Magnusson, ``Particle swarm optimization and its application to the design of diffraction grating filters," 
Opt. Lett. 32, 894-896 (2007).

\bibitem{Clerc10} J. Sun, C.-H. Lai, X.-J. Wu, {\it Particle Swarm Optimisation: Classical and Quantum Perspectives}, (CRC Press, USA,  2012).
\bibitem{Tzang18} O. Tzang, A. M. C.-Aguirre, K. Wagner and R. Piestun, ``Adaptive wavefront shaping for controlling nonlinear multimode interactions in optical fibres," 
Nat. Photon. \textbf{12}, 368–374 (2018). 

\end{thebibliography}
\end{document}